\pdfoutput=1

\documentclass[preprint, 10pt]{aastex}

\usepackage{epstopdf}
\usepackage{float}
\usepackage{subfigure}
\usepackage{lineno}

\usepackage{graphicx, subfigure}
\usepackage{float}
\graphicspath{{plots/} }

\received{June 2015}
\revised{revision date}
\accepted{acceptance date}

\shorttitle{MOLD database}

\shortauthors{Vuj\v ci\'c et al.}

\begin{document}
\title{MOL-D: A COLLISIONAL DATABASE AND WEB SERVICE WITHIN THE VIRTUAL
ATOMIC AND MOLECULAR DATA CENTER}


\author{V. Vuj\v ci\'c}
\affil{Faculty of Organizational Sciences,
    Univesity of Belgrade and Astronomical Observatory, Volgina 7, 11060 Belgrade 74,
       Serbia}

\author{D. Jevremovi\'c}
\affil{Astronomical Observatory, Volgina 7, 11060 Belgrade 74,
       Serbia}

\author{A.A.Mihajlov, Lj.M.Ignjatovi\'c, V.A.Sre\'ckovi\'c}
\affil{Institute of physics,Univesity of Belgrade,  P.O. Box 57, 11001, Belgrade, Serbia}

\email{vlada,mihajlov@ipb.ac.rs}

\author{M.S.Dimitrijevi\'c}
\affil{Astronomical Observatory, Volgina 7, 11060 Belgrade 74,
       Serbia and Observatoire de Paris, 92195 Meudon Cedex, France \\and IHIS Techno experts, Batajni\v cki put 23, 11080 Zemun, Serbia}
\email{veljko,mdimitrijevic,darko@aob.rs}

\and

\author{M. Malovi\'c}
\affil{Innovation Centre of Faculty of Technology and Metallurgy, University of Belgrade, Karnegijeva 4,  11000 Belgrade, Serbia}

\email{miodrag@malovic.in.rs}

\begin{abstract}

MOL-D database is a collection of cross-sections and rate coefficients for specific collisional
processes and a web service within the Serbian Virtual Observatory (SerVO \footnote{http://servo.aob.rs})
and the Virtual Atomic and Molecular Data Center (VAMDC \footnote{http://www.vamdc.eu}).
This database contains photo-dissociation cross-sections for the individual ro-vibrational states
of the diatomic molecular ions and rate coefficients for the atom-Rydberg atom chemi-ionization
and inverse electron-ion-atom chemi-recombination processes. At the moment it contains
data for photodissociation cross-sections of hydrogen  H$_2^+$ and helium He$_2^+$ molecular
ions and the corresponding averaged thermal photodissociation cross-sections. The ro-vibrational energy states
and the corresponding dipole matrix elements are provided as well.
Hydrogen and helium molecular ion  data are  important for calculation of solar and stellar atmosphere models
and for radiative transport,  as well as for kinetics
of other astrophysical and laboratory plasma (i.e. early Universe).

\end{abstract}


\section{Introduction}

Mihajlov and coworkers \citep{mih86,mih93,mih07,ign09,ign14b,sre14}
have demonstrated that ion-atom radiative processes, the photodissociation of the diatomic
molecular ion in the symmetric and non-symmetric cases, could be important in
specific
stellar atmosphere layers and they should be  included in chemical models.
In the symmetric case, we consider the processes of molecular ion
photodissociation (bound-free) and ion-atom photoassociation (free-bound):
\begin{equation}
\label{eq:sim1} h\nu + A_{2}^{+}  \Longleftrightarrow A + A^{+},
\end{equation}
and the corresponding free-free absorption and emission processes:
\begin{equation}
\label{eq:sim2} h\nu + A + A^{+}  \Longleftrightarrow  A^{+} +A,
\end{equation}
where $A$ and $A^{+}$ are atom and ion in their ground states, and
$A_{2}^{+}$ is molecular-ion in the ground electronic state.

In the non-symmetric case, the similar processes
of photodissociation/photoassociation are:
\begin{equation}
\label{eq:nsim1} h\nu + AX^{+}  \Longleftrightarrow A^{+} + X,
\end{equation}
and the corresponding absorption/emission processes
\begin{equation}
\label{eq:nsim2} h\nu + A + X^{+}  \Longleftrightarrow  A^{+} +X.
\end{equation}
\noindent The processes of stimulated photoassociation, characteristic for
the non-symmetric case are:
\begin{equation}
\label{eq:nsim3} \varepsilon_{\lambda}
+ A + X^{+} \Longleftrightarrow (AX^{+})^{*} ,
\end{equation}
where $X$ is an atom whose ionization potential is less than the corresponding value for atom $A$.
$AX^{+}$ is also molecular-ion in the ground electronic state and $(AX^{+})^{*}$
molecular-ion in the first excited electronic state.

In the general case molecular ion $A_{2}^{+}$ or $AX^{+}$ can be in
one of the states from the group which contains the ground electronic state. Similarly,
the excited molecular ion $(AX^{+})^{*}$  can exist in one of the states
from the group which contains the first excited electronic state of the
considered molecular ion.

For the solar atmosphere,  $A$ usually denotes atom H(1s) and $X$ one of the
relevant
metal atoms (Mg, Si, Ca, Na ...) \citep{mih86,mih93,mih07,ign14b,sre14}, but
there  are cases where
$A$=He, and $X$=H, Mg, Si, Ca, Na. For the helium-rich white dwarf
atmospheres $A$ denotes He(1s$^2$) and $X$ denotes, H(1s), and eventually C, O. \citep{mih92,mih13,ign14a}.

Our results show the importance of including the symmetric processes with $A=$H(1s)
in the stellar atmosphere models (see e.g. \citet{fon09})
and for the early Universe investigation (see \citet{cop13}).
Also, our results for $A=$He$(1s^{2})$ have been used
for modeling the DB white dwarf atmospheres (Koester, private communication).
Such data are also of interest for research on the corresponding
weakly ionized laboratory plasmas.

The processes mentioned above are closely connected to several groups
of inelastic atomic collision processes. The first few groups consist of the
chemi-ionization
processes  in symmetric and non-symmetric atom/Rydberg-atom collisions, including the processes
of associative ionization as well as the corresponding inverse electron-ion-atom
chemi-recombination processes:
\begin{equation}
\label{eq:ch1} A^{*}(n) + A  \Longleftrightarrow   A_{2}^{+} + e,
\end{equation}
\begin{equation}
\label{eq:ch2} A^{*}(n) + A  \Longleftrightarrow A + A^{+} + e,
\end{equation}
\begin{equation}
\label{eq:ch3} A^{*}(n) + X  \Longleftrightarrow   A_{2}^{+} + e,
\end{equation}
\begin{equation}
\label{eq:ch4} A^{*}(n) + X  \Longleftrightarrow A + X^{+} + e.
\end{equation}
\noindent $A^{*}(n)$ is an atom in one of the highly excited (Rydberg) states
with the principal quantum number $n \gg 1$, $e$ is a free electron and
$A$, $A^{+}$ $A_{2}^{+}$, $X$ have the same meaning as in the previous cases.
The ionization potential of the atom X is lower
than that of the atom A.
The considered radiative processes are allowed by the
dipole selection rules.

\begin{figure}[h!]
\centerline{\includegraphics[width=\columnwidth, height=0.75\columnwidth]{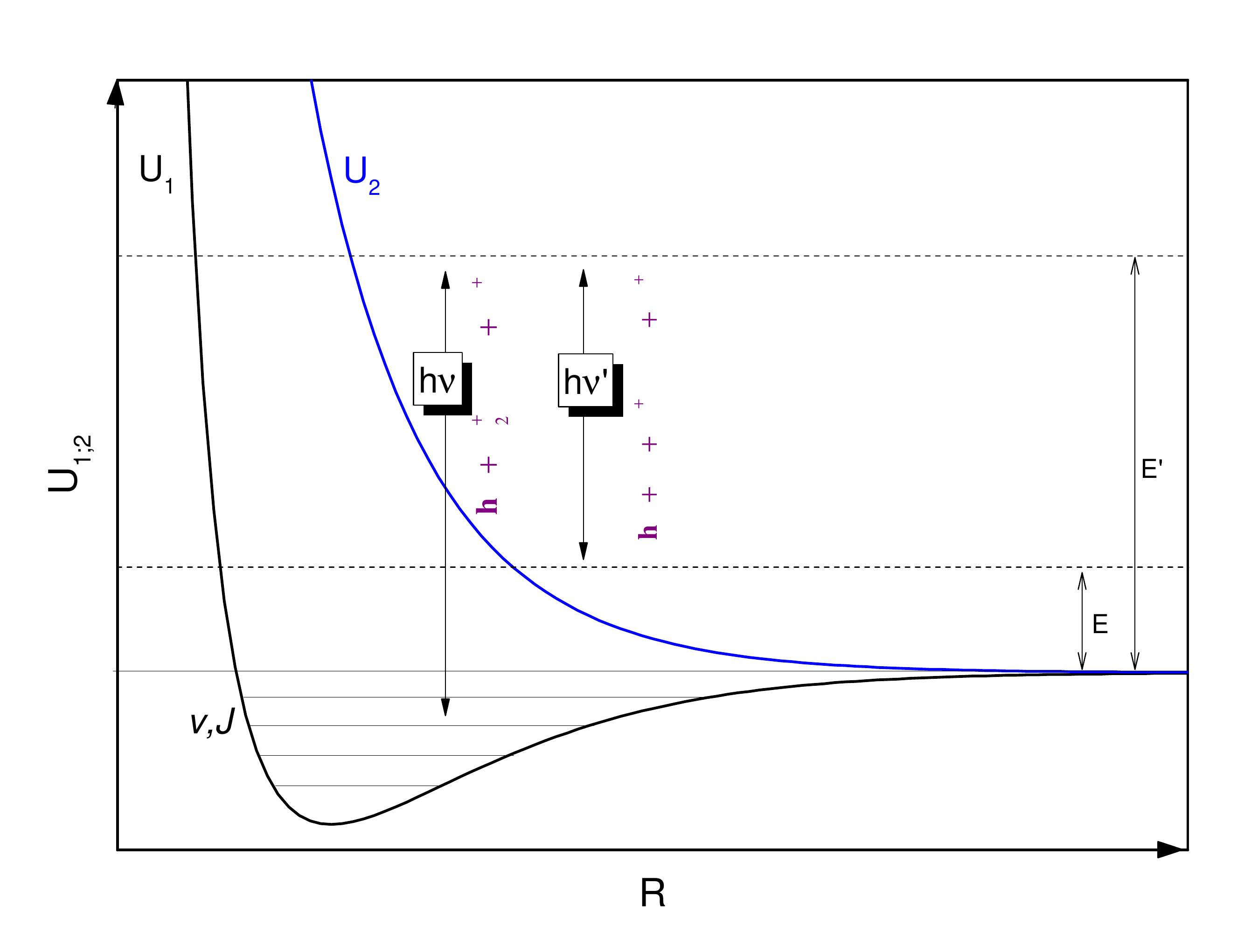}}
\caption{The schematic presentation of the photo-dissociation/association processes [Eq. (\ref{eq:sim1})]
and free-free processes [Eq. (\ref{eq:sim2})]: R is the internuclear distance,
U$_1$(R) and U$_2$(R) are the potential energy curves of the initial(lower) and final(upper) electronic state of molecular ion $A_{2}^{+}$, $J$ and $v$ are
individual ro-vibrational state, $E$ and $E'$ are the total energies of the system $A+A^{+}$, $h\nu$ and $h\nu'$ are the photon energies.} \label{fig:SHEMA1}
\end{figure}
\begin{figure}
\centerline{\includegraphics[width=\columnwidth, height=0.75\columnwidth]{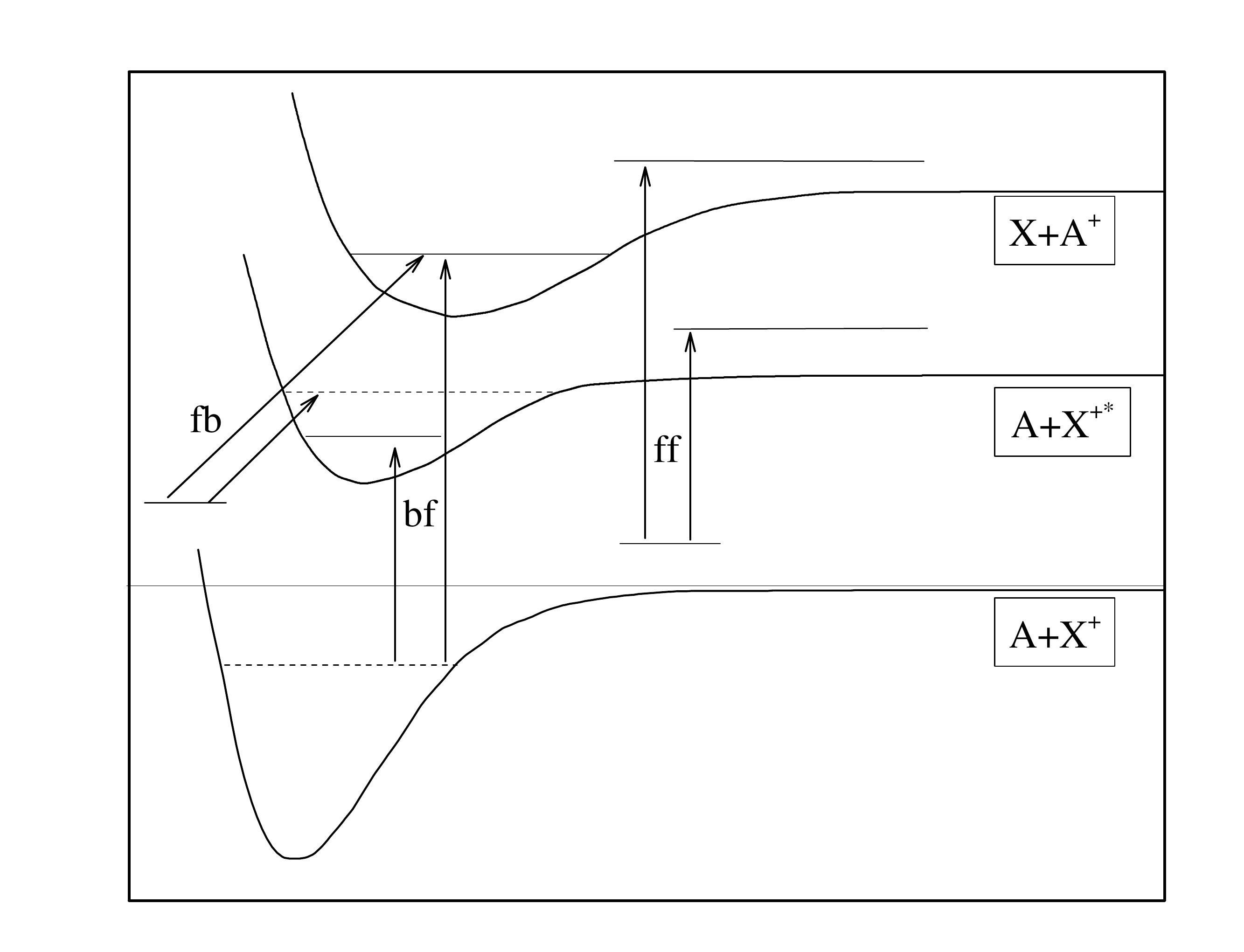}}
\caption{The schematic presentation of the processes [Eq. (\ref{eq:nsim1})]
-[Eq. (\ref{eq:nsim3})] for the case of the molecular ion $AX^{+}$ and ion-atom system $A^{+}+X$} \label{fig:SHEMA2}
\end{figure}
\begin{figure}
\centerline{\includegraphics[width=\columnwidth, height=0.75\columnwidth]{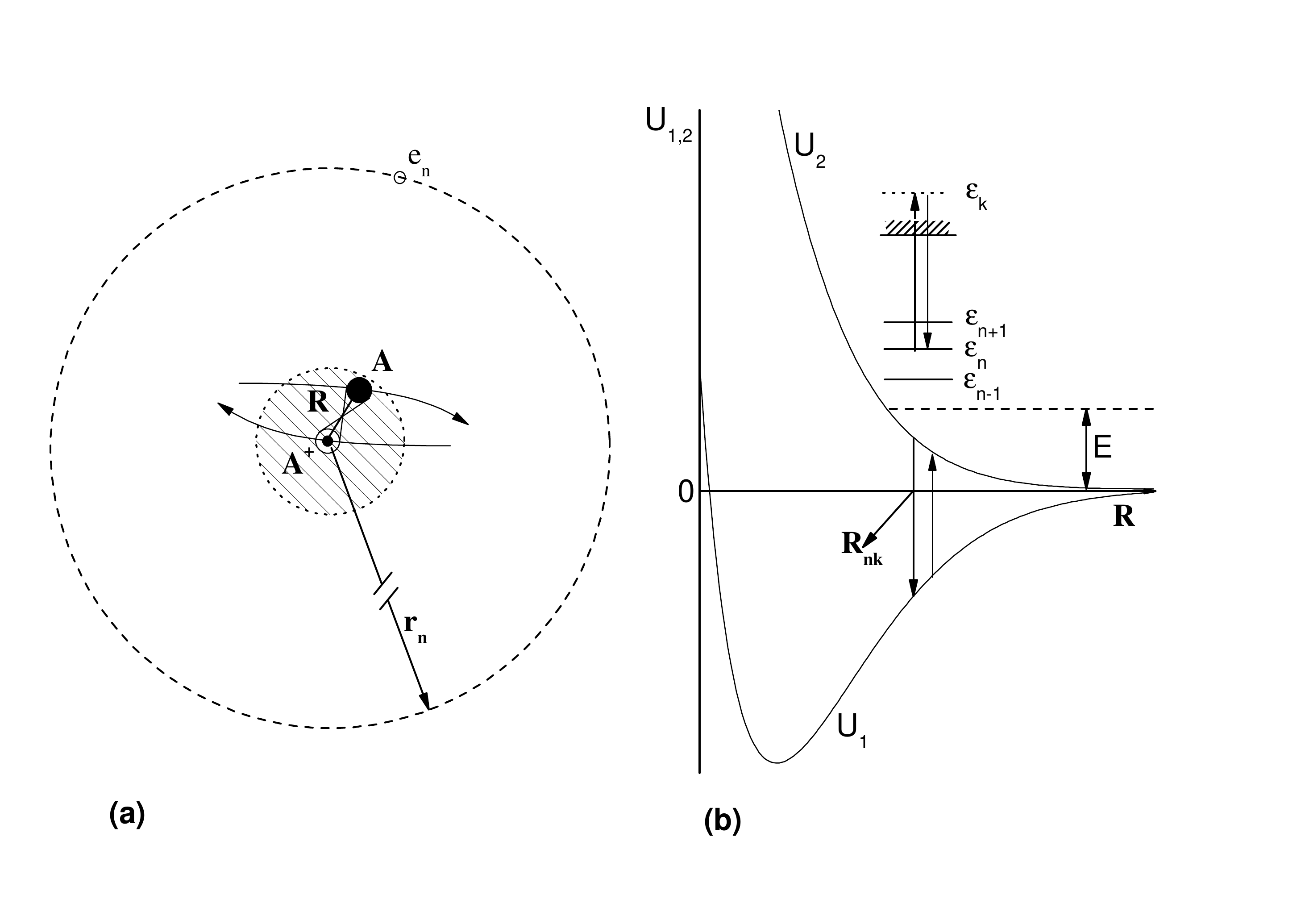}}
\caption{The schematic presentation of the chemi-ionization/recombination processes [Eqs.(\ref{eq:ch1}-\ref{eq:ch4})]:
n is the principal quantum number of the Rydberg state, $\epsilon_{k}$ is the energy of the free electron.} \label{fig:SHEMA3}
\end{figure}
\begin{figure}
\centerline{\includegraphics[width=\columnwidth, height=0.75\columnwidth]{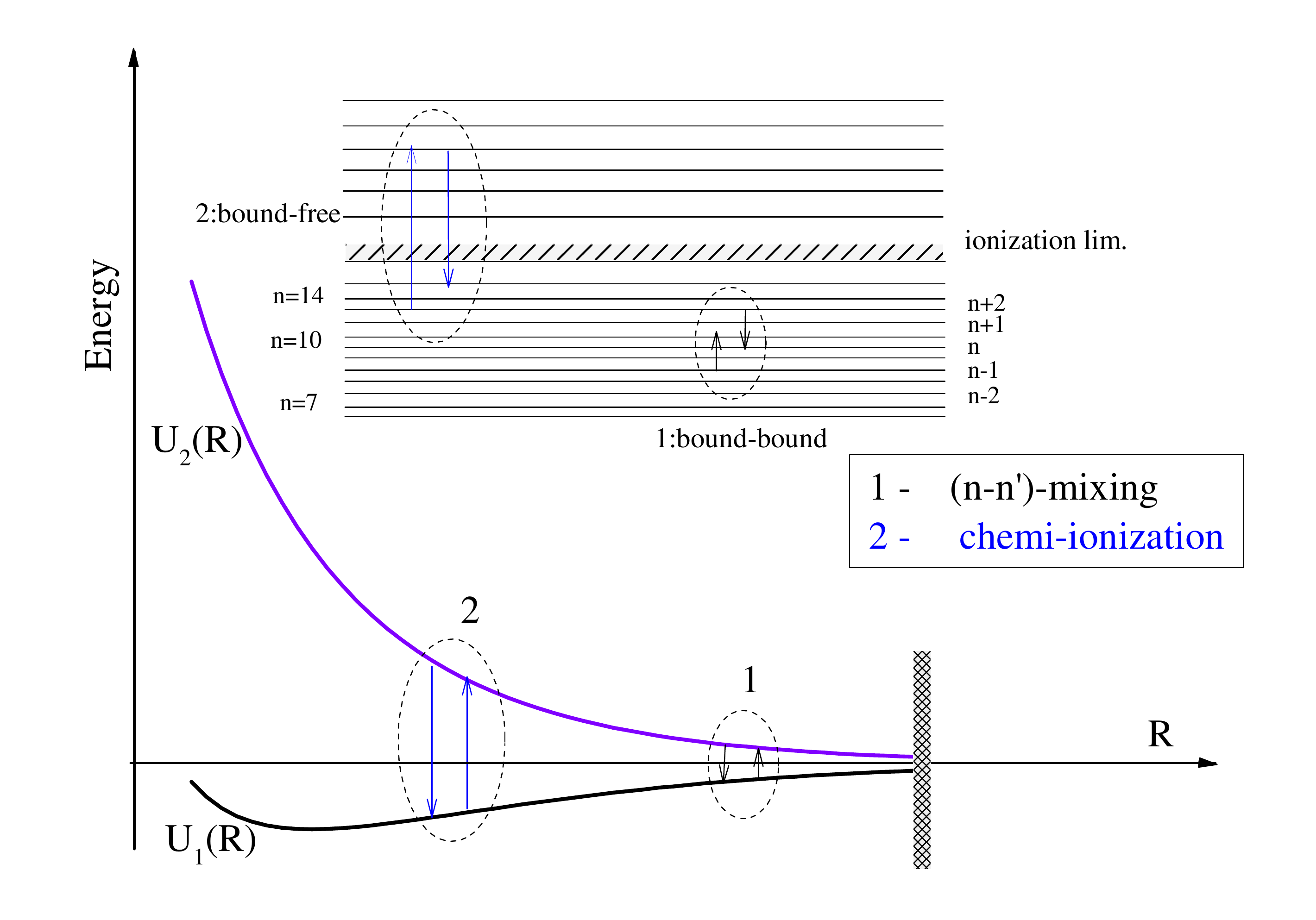}}
\caption{The schematic presentation of the chemi-ionization and $(n-n')$-mixing processes [Eqs. (\ref{eq:ch2} and \ref{eq:mix1})]:
n and n' are the principal quantum numbers of the considered Rydberg states.} \label{fig:SHEMA4}
\end{figure}

The other groups of processes consist of excitation and deexcitation processes known also as
the $(n-n')$-mixing processes:
\begin{equation}
\label{eq:mix1}
A^{*}(n)+A \Longrightarrow \left\lbrace
\begin{array}{ll}
\displaystyle{ A^{*}(n')+A,}\\
\displaystyle{ \qquad \qquad \qquad  1 \ll n< n',}\\
\displaystyle{ A+A^{*}(n'),}
\end{array}
\right.
\end{equation}
and
\begin{equation}
\label{eq:mix2}
A^{*}(n)+A \Longrightarrow \left\lbrace
\begin{array}{ll}
\displaystyle{ A^{*}(n')+A,}\\
\displaystyle{ \qquad \qquad \qquad  1 \ll n'< n}\\
\displaystyle{ A+A^{*}(n'),}
\end{array}
\right.
\end{equation}
\noindent $A^{*}$ has the same meaning as in the case of
chemi-ionization/chemi-recombination processes.

Chemi-ionization/chemi-recombination and
$(n-n')$-mixing processes are such that system passes through the phase
where it can be treated in the form:
\begin{equation}
\label{eq:qm1} (A+A^{+})_{q.-m.}^{in,fin} + e_{out},
\end{equation}
or
\begin{equation}
\label{eq:qm2} (A+X^{+})_{q.-m.}^{in,fin} + e_{out}.
\end{equation}
\noindent ($A+A^{+}$)$_{q.-m.}^{in,fin}$ and ($A+X^{+}$)$_{q.-m.}^{in,fin}$ denote a quasi-molecular
ion-atom complex in the corresponding (initial or final) electronic state, and $e_{out}$
denotes a free electron in weakly bound or free state.

The  connection of these processes with the above described radiative processes
is in the following chemi-ionization/chemi-recombination
and $(n-n')$-mixing transitions:

\begin{equation}
\label{eq:qm3} (A+A^{+})_{q.-m.}^{in} \longrightarrow (A+A^{+})_{q.-m.}^{fin}
\end{equation}
or
\begin{equation}
\label{eq:qm4} (A+X^{+})_{q.-m.}^{in} \longrightarrow (A+X^{+})_{q.-m.}^{fin}.
\end{equation}


The processes described above (Eqs. (1-11)) are  schematically illustrated in Figs. (1-6).

The results obtained during the investigation of the processes mentioned in the present section
are presented  in MOL-D database which will be described in the next section.
The first version of this database is available online and can be accessed directly through
http://servo.aob.rs/mold or through VAMDC node within the
Serbian Virtual Observatory (SerVO \citet{jev09}), http://servo.aob.rs
and the Virtual Atomic and Molecular Data Center
(VAMDC \citet{dub10,rix11}, http://www.vamdc.org).

\section{Content of MOL-D e-service}

The MOL-D is an e-service which exposes our results to the wider community.
In particular, we provide:

\begin{itemize}
  \item the cross-sections for the photodissociation of individual ro-vibrational states of
  the considered molecular ions as well as the cross sections for the inverse ion-atom photoassociation,
  (Eq.(\ref{eq:sim1}) and Eq. (\ref{eq:nsim1})), in a wide region of photon wavelengths $\lambda$,

  \item the averaged thermal cross sections for the considered molecular ion photodissociation and
  for the reverse process, ion-atom photoassociation, (Eq.(\ref{eq:sim1}) and Eq. (\ref{eq:nsim1})),
  in wide region of $\lambda$ and temperature $T$,

  \item the rate coefficients and the corresponding averaged thermal cross sections
  for the ion-atom absorption processes and inverse emission processes, Eq. (\ref{eq:sim2})
  and Eq. (\ref{eq:nsim2}), as well as for the non-symmetric ion-atom photoassociation
  in Eq. (\ref{eq:nsim3}) in wide region of $\lambda$ and $T$,



  \item visualization of the wavelength dependance of the averaged  thermal cross
    sections for  a given temperature input.

i
\end{itemize}

MOL-D is available online from the end of 2014 and for the moment it contains the data for the
photodissociation processes Eq. (\ref{eq:sim1}) with $A=$ H(1s)and $A=$ He(1s$^2$). In the near future
we intend  to include the relevant data for some other non-symmetric photodissociation processes
Eq. (\ref{eq:nsim1}).

The cross-sections for the photodissociation of individual ro-vibrational state of the considered molecular
ions are determined in the dipole approximation:
\begin{equation}
\sigma_{J,v}(\lambda)  =  \frac{8\pi^{3}}{3\lambda}\left[ \frac{(J+1)|D_{E,J+1;v,J}|^{2}+J|D_{E,J-1;v,J}|^{2}}{2J+1}
\right],\\
\end{equation}

\noindent and the corresponding averaged thermal cross sections  are given by:

\begin{equation}
\sigma_{\rm ph}(\lambda,T)  =  \frac{1}{Z} \sum_J
\sum_{v} g_{J;v}(2J+1)
e^{-\frac{E_{Jv}-E_{00}}{k_{\rm B}T}}\sigma_{J,v}(\lambda).
\end{equation}

\noindent $D_{E,J+1;v,J}$ is the relevant dipole matrix element, $E_{Jv}$ is the energy of the individual states
with the angular and vibrational quantum numbers $J$ and $v$ respectively, and $Z$ is the partition function
\begin{equation}
Z  =\sum_J  \sum_{v} g_{J;v}(2J+1)
e^{-\frac{E_{Jv}-E_{0,0}}{k_{\rm B}T}}.
\end{equation}
In this expression the product $g_{J;v}\cdot(2J+1)$ is the statistical weight of the
considered state and the coefficient $g_{J;v}$ depends on the "the spin of the nuclei".

We also plan to include  the rate coefficients for the chemi-ionization/recombination
and $(n-n')$-mixing, (Eqs. (\ref{eq:ch1})-(\ref{eq:mix2})).
The values of the rate coefficients will be determined in the semi-classical approach
(see e.g. \citet{mih11}), but using a significantly improved numerical procedure respect to
the previous papers \citep{mih03,mih04,mih05,mih08,mih11,sre13}.

\section{Technical characteristics of MOL-D database}
\label{sec:tech}

\begin{figure}[h!]
\centerline{\includegraphics[width=\columnwidth, height=0.75\columnwidth]{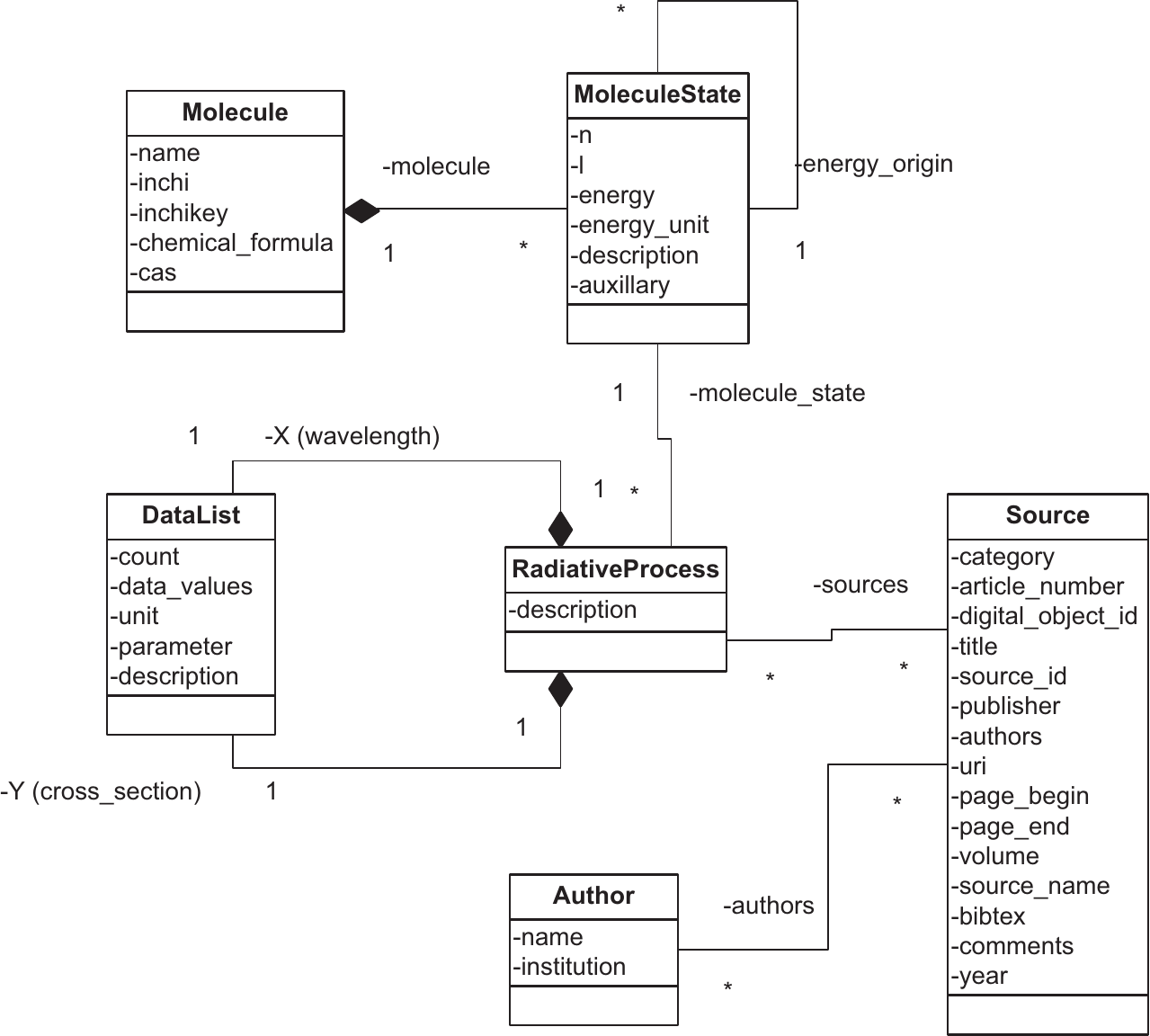}}
\caption{Static structure of the MOL-D database. Relationships between entities are shown by connected lines with designated cardinalities ("1" and "*" denote one and many, respectively), i.e a molecule can have multiple states.} \label{fig:SHEMA5}
\end{figure}

The principal structure of the Belgrade MOL-D database is shown schematically
in Fig. \ref{fig:SHEMA5} using UML notation.

MOL-D data are exposed as a web form and a web service accesible according to
VAMDC specification
\footnote{http://vamdc-standards.readthedocs.org/en/latest/dataAccessProtocol/vamdctap.html}. VAMDC
\citep{dub10} standards
define VAMDC-TAP RESTful web-service request using VSS2 query language and results formatted as a XSAMS
(XML Schema for Atoms, Molecules and Solids) document \footnote{http://vamdc.eu/documents/standards/dataModel/vamdcxsams/index.html}. Such approach enables accessing multiple databases in a single query.

Software is built on top of Django, a web-based Python framework, and represents an adaptation and extension
of VAMDC NodeSoftware. User interface is AJAX-enabled, using JQuery javascript framework and plots are generated by pyplot (matplotlib).

\begin{figure*}
\center{\includegraphics[
scale=0.4]{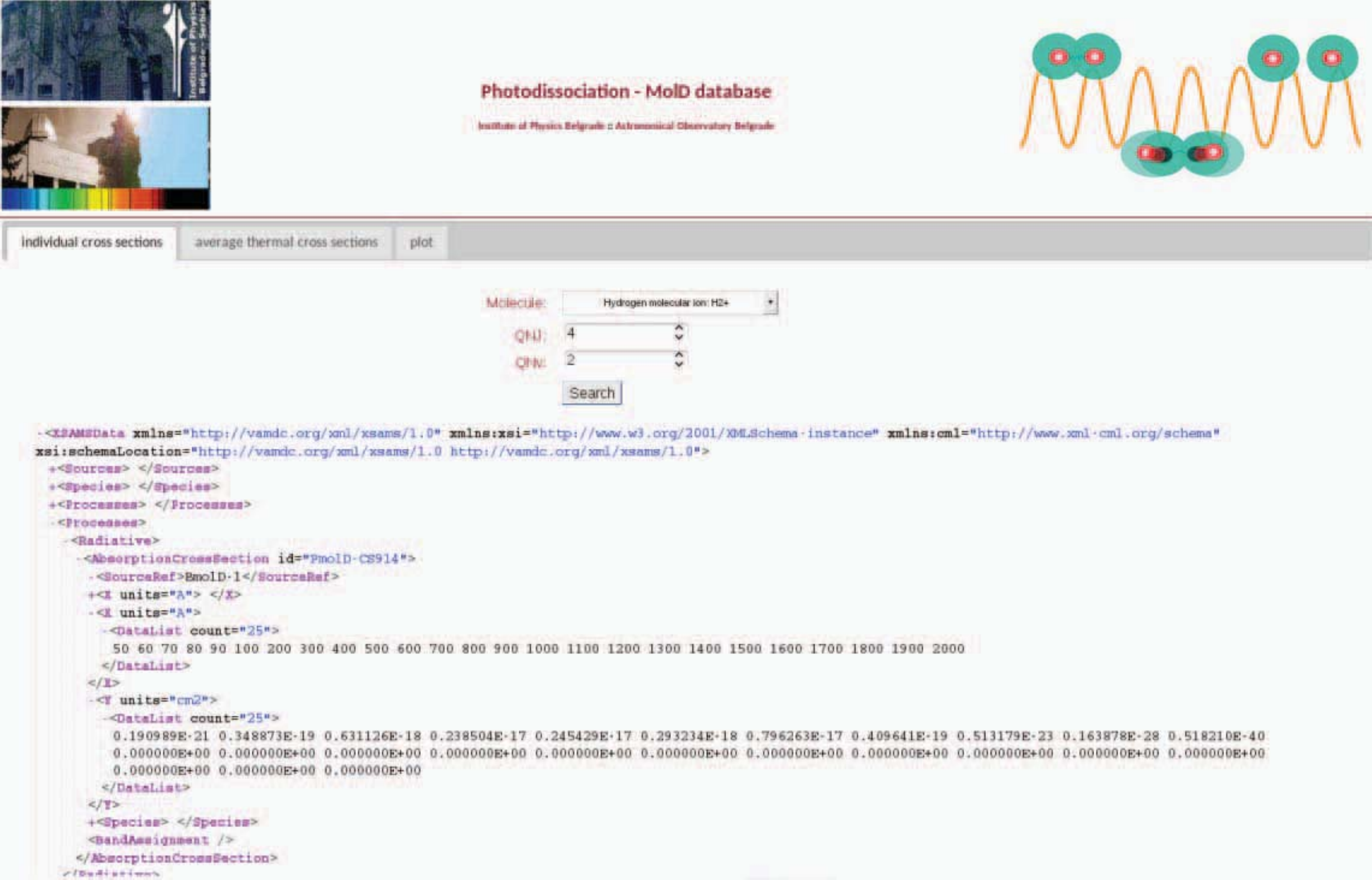}

\vspace{1cm}

\includegraphics[scale=0.4]
{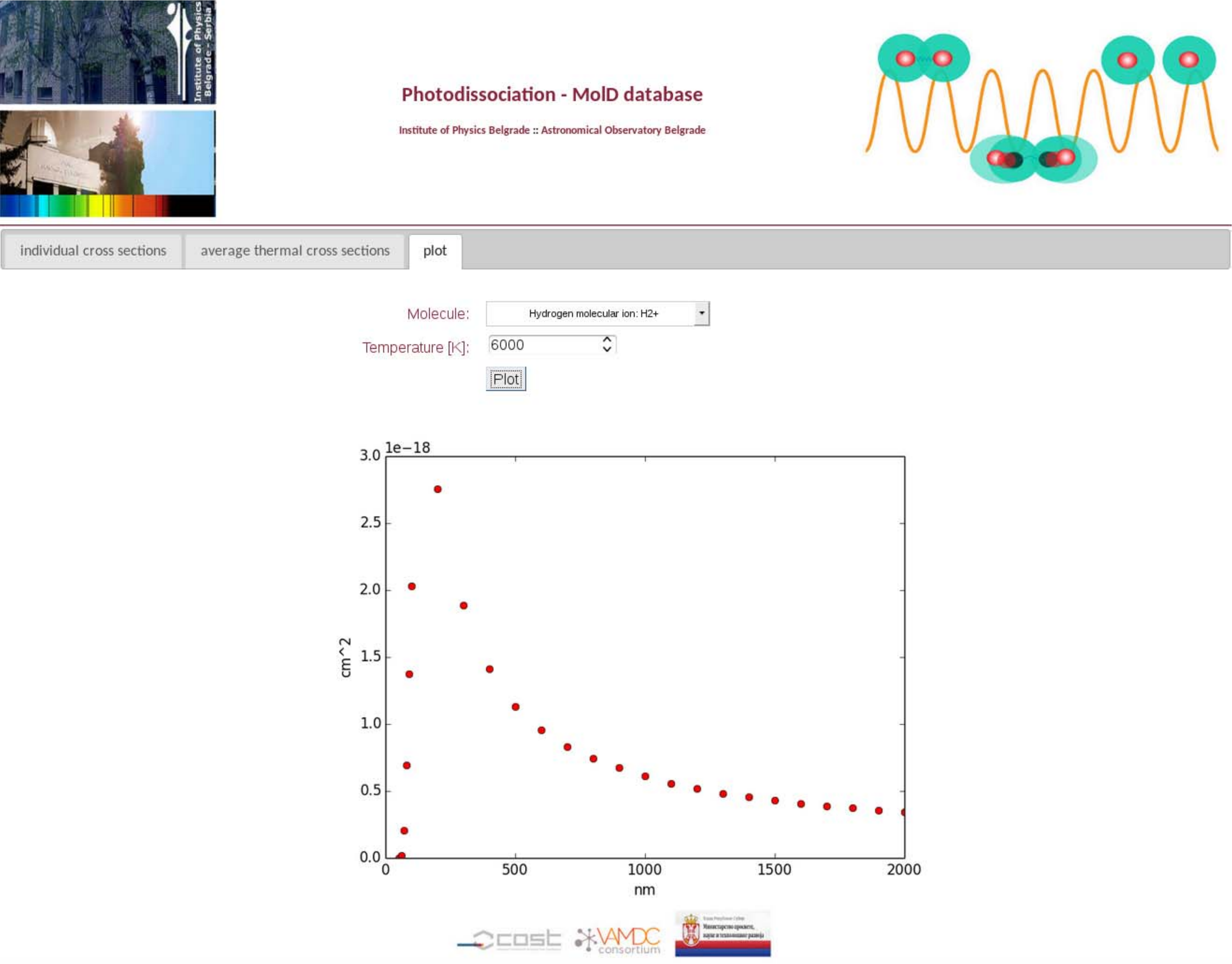}}

 \caption{Some screen shots of the MOL-D node at the Belgrade server station.}
\label{fig:shema6}
\end{figure*}

Besides acting as a VAMDC-compliant web service, accessible through VAMDC portal and other tools implemented on VAMDC standards,
MOL-D offers on-site services:
\begin{itemize}
  \item {user can make a selection based on molecule and quantum number J (QNJ) or quantum number v (QNv)}
  \item {calculate averaged thermal cross section based on the temperature for a specific molecule and wavelength}
  \item {make a plot of averaged thermal cross sections along all (discrete) wavelengths for a given temperature}
\end{itemize}
A screen shot of MOL-D Database at Belgrade server station \footnote{http://servo.aob.rs/mold} is
shown in Fig. \ref{fig:shema6}. An example of the visualization of a data set that represents the
averaged cross section versus wavelength is shown in the right panel of Fig. \ref{fig:shema6}.

\section{FUTURE DEVELOPMENT AND PERSPECTIVES}
\label{sec:fut}

In the near future we plan to add the results
of  the rate coefficients for the ion-atom absorption processes
and inverse emission processes. We will also include data for the non-symmetric ion-atom photoassociation. Our plans also consist of including the rate coefficients
for chemi-ionization in atom-Rydberg atom collisions (including the processes
of associative and Penning type ionization) and corresponding
inverse chemi-recombination processes in electron-ion-atom collisions.
Finally, we intend  to include  the rate coefficients for the excitation and deexcitation
$(n-n')$-mixing processes in the relevant region of the principal quantum number $n$ and $T$.


 We plan to calculate and include new data about processes which involve
species such as HeH$^{+}$, LiH$^{+}$, NaH$^{+}$, SiH$^{+}$ which are important
for the early Universe chemistry and for the modeling of stellar and solar atmospheres. The MOL-D database  will be regularly updated with new results.

\acknowledgments

The authors are thankful to the Ministry of Education, Science and Technological Development
of the Republic of Serbia for the support of this work within the projects 176002 and III44002.
A part of this work has been supported by VAMDC. VAMDC is funded under the Combination of
Collaborative Projects and Coordination and Support Actions Funding
Scheme of The Seventh Framework Program. We are grateful to  Dr Guy Rixon for his help in all aspects of inclusion of MOLD database in VAMDC.


\begin{thebibliography}{31}
\expandafter\ifx\csname
natexlab\endcsname\relax\def\natexlab#1{#1}\fi


\bibitem[Coppola et al.(2013)]{cop13} Coppola, C.~M., Galli,
D., Palla, F., Longo, S., \& Chluba, J.\ 2013, \mnras, 434, 114

\bibitem[Dubernet et al.(2010)]{dub10} Dubernet, M.~L.,
Boudon, V., Culhane, J.~L., et al.\ 2010, \jqsrt, 111, 2151

\bibitem[Fontenla et al.(2009)]{fon09} Fontenla, J.~M.,
Curdt, W., Haberreiter, M., Harder, J., \& Tian, H.\ 2009, \apj, 707, 482

\bibitem[Ignjatovi{\'c} et al.(2009)]{ign09} Ignjatovi{\'c},
L.~M., Mihajlov, A.~A., Sakan, N.~M., Dimitrijevi{\'c}, M.~S.,
\& Metropoulos, A.\ 2009, \mnras, 396, 2201

\bibitem[Ignjatovi{\'c} et al.(2014a)]{ign14a} Ignjatovi{\'c},
L.~M., Mihajlov, A.~A., Sre{\'c}kovi{\'c}, V.~A.,
\& Dimitrijevi{\'c}, M.~S.\ 2014a, \mnras, 441, 1504

\bibitem[Ignjatovi{\'c} et al.(2014b)]{ign14b} Ignjatovi{\'c},
L.~M., Mihajlov, A.~A., Sre{\'c}kovi{\'c}, V.~A.,
\& Dimitrijevi{\'c}, M.~S.\ 2014b, \mnras, 439, 2342

\bibitem[Jevremovi{\'c} et al.(2009)]{jev09} Jevremovi{\'c},
D., Dimitrijevi{\'c}, M.~S., Popovi{\'c}, L.~{\v C}., et al.\ 2009, \nar,
53, 222

\bibitem[Koester (2015)]{koe2015} Koester,
D., Private comunication \ 2015

\bibitem[Mihajlov
\& Dimitrijevic(1986)]{mih86} Mihajlov, A.~A., \& Dimitrijevic, M.~S.\ 1986, \aap, 155, 319

\bibitem[Mihajlov
\& Dimitrijevic(1992)]{mih92} Mihajlov, A.~A., \& Dimitrijevic, M.~S.\ 1992, \aap, 256, 305

\bibitem[Mihajlov et al.(1993)]{mih93} Mihajlov, A.~A., Dimitrijevic, M.~S., \& Ignjatovic, L.~M.\ 1993, \aap, 276, 187

\bibitem[Mihajlov et al.(2013)]{mih13} Mihajlov, A.~A.,
Ignjatovi{\'c}, L.~M., Sre{\'c}kovi{\'c}, V.~A., Dimitrijevi{\'c}, M.~S.,
\& Metropoulos, A.\ 2013, \mnras, 431, 589

\bibitem[Mihajlov et al.(2007)]{mih07} Mihajlov, A.~A., Ignjatovi{\'c}, L.~M.,
Sakan, N.~M., \& Dimitrijevi{\'c}, M.~S.\ 2007, \aap, 469, 749

\bibitem[Mihajlov et al.(2011)]{mih11} Mihajlov, A.~A.,
Ignjatovi{\'c}, L.~M., Sre{\'c}kovi{\'c}, V.~A.,
\& Dimitrijevi{\'c}, M.~S.\ 2011, \apjs, 193, 2

\bibitem[Mihajlov et al.(2003)]{mih03} Mihajlov, A.~A.,
Ignjatovi{\'c}, L.~M., Dimitrijevi{\'c}, M.~S.,
\& Djuri{\'c}, Z.\ 2003, \apjs, 147, 369

\bibitem[Mihajlov et
al.(2005)]{mih05} Mihajlov, A.~A., Ignjatovi{\'c}, L.~M., \& Dimitrijevi{\'c}, M.~S.\ 2005, \aap, 437, 1023

\bibitem[Mihajlov et al.(2004)]{mih04} Mihajlov, A.~A.,
Ignjatovic, L.~M., Djuric, Z.,
\& Ljepojevic, N.~N.\ 2004, JPhys B Atomic Molecular Physics, 37, 4493

\bibitem[Mihajlov et al.(2008)]{mih08} Mihajlov, A.~A.,
Ignjatovi{\'c}, L.~M., Sre{\'c}kovi{\'c}, V.~A.,
\& Djuri{\'c}, Z.\ 2008, \jqsrt, 109, 853


\bibitem[Rixon et al.(2011)]{rix11} Rixon, G., Dubernet,
M.~L., Piskunov, N., et al.\ 2011, American Institute of Physics Conference
Series, 1344, 107

\bibitem[Sre{\'c}kovi{\'c} et al.(2014)]{sre14}
Sre{\'c}kovi{\'c}, V.~A., Mihajlov, A.~A., Ignjatovi{\'c}, L.~M.,
\& Dimitrijevi{\'c}, M.~S.\ 2014, Adv.Sp.Res., 54, 1264


\bibitem[Sre{\'c}kovi{\'c} et
al.(2013)]{sre13} Sre{\'c}kovi{\'c}, V.~A., Mihajlov, A.~A., Ignjatovi{\'c}, L.~M., \& Dimitrijevi{\'c}, M.~S.\ 2013, \aap, 552, AA33

%

%

\end{thebibliography}

\newcommand{\noopsort}[1]{} \newcommand{\printfirst}[2]{#1}
  \newcommand{\singleletter}[1]{#1} \newcommand{\switchargs}[2]{#2#1}

\end{document}